\documentclass[twocolumn,showpacs,preprintnumbers,amsmath,amssymb]{revtex4}

\begin{document}
\title{Holographic Bound on Information in Inflationary Perturbations}
\author{Craig J. Hogan}
\address{Astronomy and Physics Departments, 
University of Washington,
Seattle, Washington 98195-1580}
\begin{abstract}
The formation  of frozen classical perturbations from vacuum quantum fluctuations during inflation  is described as a unitary quantum process with  apparent ``decoherence" caused by the expanding spacetime. 
It is argued that the maximum observable information content per comoving volume in classical modes  is subject to the covariant entropy bound at the  time those modes decohere, leading to  a new quantitative bound  on the  information contained in frozen field modes in  phase space.  This bound implies holographic correlations of large-scale cosmological perturbations  that may be observable.
\end{abstract}
\pacs{98.80.-k}
\maketitle

\section{Introduction}
The mostly widely held theory of cosmological structure  is based on primordial fluctuations that originate from quantum fields  during inflation.  The expansion of spacetime  converts virtual quanta--- the zero-point vacuum fluctuations of fields--- into real classical field perturbations. 
Inflation theory\cite{lythriotto,liddlelyth,Starobinsky:1979ty,Hawking:1982cz,Guth:1982ec,Bardeen:1983qw,Starobinsky:1982ee,Halliwell:1985eu,guthpi,Grishchuk:1989ss,Grishchuk:1990bj,Albrecht:1994kf,Lesgourgues:1997jc,Polarski:1996jg,kiefer,Kiefer:1999sj} describes  how the quantum state of each field mode changes character as it expands to exceed the size of the apparent horizon---  from eigenstates of  number (in particular,  an initial vacuum state  with zero particles) to eigenstates  of field amplitude,  in which    the quantum zero-point field fluctuations are frozen as real quasi-classical observables. 
Recent data, especially the concordance of microwave background anisotropy\cite{Bennett:2003bz,Komatsu:2003fd,Spergel:2003cb} and galaxy clustering\cite{Tegmark:2003uf}, confirm many detailed features of this basic picture, including the primordial origin well before recombination,  a nearly scale-invariant power spectrum,  and approximately Gaussian statistics. 
It has long been
hoped that detailed study of these quantum structures, dating as they do from close to the Planck time, might reveal qualitatively new  fundamental physics connected with quantum gravity (see e.g. \cite{Kempf:1998gk,Niemeyer:2000eh,Kempf:2000ac,Kempf:2001fa,brandenberger01,martin01,starobinsky01,easther,hui,Albrecht:2002xs,Keski-Vakkuri:2003vj}). 

A radical but concretely formulated conjecture  about such physics, based on considerations such as quantum unitarity during black hole evaporation, on analysis of certain systems such as extremal black holes, and on  the AdS/CFT duality, is  that nature imposes a holographic bound   on the total entropy of systems\cite{'tHooft:1999bw,'tHooft:1985re,Susskind:1993if,susskind95,bousso02}. According to this conjecture, the maximum entropy of a compact system  is much less than in standard field theory.

The  detailed pattern of cosmic perturbations preserves more than just the power spectrum: on the largest scales, it directly records  the detailed spatial configuration of the  original  field fluctuations frozen in during  inflation. 
The process of freezing out creates spatially localized information from quantum states close to the Planck time  that survives to the present.
 It is shown here that
 the entropy bound on fields  during inflation limits the amount of information eventually carried by the final classical perturbations. Although this effect leaves the predicted mean power spectrum and Gaussian amplitude distribution unchanged, it implies  a major qualitative difference from standard inflation, whose  random phases and  continuous spectrum contain in principle an infinite amount of information. A quantitative upper bound is derived here on  the mean density   of observable  information in  classical perturbation modes. This feature implies new  correlations  among modes not predicted in standard field theory.  The effect can in principle provide a direct observational test of the holographic conjecture and a probe of how it is implemented in nature.

\section{Freezing of Inflationary Quanta}

It is useful to recall the relationship of inflationary quantum field states with classically observed mode amplitudes (see e.g. \cite{guthpi,Grishchuk:1989ss,Grishchuk:1990bj,Albrecht:1994kf,Lesgourgues:1997jc,Polarski:1996jg,kiefer, Kiefer:1999sj}). 

The final  eigenstates correspond to definite values of field amplitude $u$. A mode that starts off in the vacuum eigenstate  at early times   ends up  as a superposition of these field-amplitude eigenstates at late times, with  a Gaussian distribution of coefficients.

We do not observe this superposition, but only one of the amplitude eigenstates. (More accurately, we observe  the late-time effects of a spacetime metric perturbation coherently imprinted by the field amplitude in one of these states). The von Neumann description  of quantum measurement   says that the wavefunction   collapses into an eigenstate when it becomes classical. The more modern view is that the whole linearly-evolving wavefunction never collapses; however, decoherence causes the entire macroscopic world to correlate with only one of the eigenstate outcomes in such a way that the other branches of the wavefunction are unobservable.\cite{Gell-Mann:1995cu,Hartle:2002nq}
 For a given inflationary mode of comoving wavenumber $k$, this apparent decoherence occurs near the time $t_k$  when  $k= aH$, where $a(t)$ is the cosmic scale factor and $H$ denotes the expansion rate during inflation.  The freezing does not depend on  observations, but is a natural process during inflation as a mode's wavelength expands beyond the apparent horizon.

Any observation yields one of the eigenvalues $u$ (with probability  given by the standard Gaussian amplitude distribution.) The different $u$ are macroscopically distinguishable options like Schr\"odinger's live and dead cats, on a grand scale:  they correspond eventually to entirely different distributions of galaxies.  The information corresponding to the superposition of eigenstates  is contained in observable correlations.  The information content  of  large scale classical observables associated with modes on scale $k$ is  subject to the bound  on  total information content at the time $t_k$.
 
The bound derived here is based on the assumption that although the entropy of the universe today is vastly greater than that of the same comoving volume during inflation, the ``frozen'' quantum fluctuations on large scales, and the metric perturbations that arise from them, are subject to the holographic entropy bounds at the time they freeze out.

\section{Holographic bound on frozen information}

It has been conjectured that  nature  imposes a fundamental limit on total entropy that applies to all fields. A general formulation of this limit, to which no exceptions have been found,  is the ``covariant entropy bound''\cite{bousso02}. Consider a closed spacelike 2-surface. Construct a    null 3-volume  $\cal{V}$ by propagating null rays inwards from the surface,    into the future and the past, such that the areas of the inward-propagating light fronts are everywhere decreasing. The covariant entropy bound states that the entropy of $\cal{V}$ on either the future or past surfaces does not exceed one quarter of  the area of the bounding 2-surface in Planck units.

 In the inflationary context, the largest 2-sphere allowing this construction has radius   just slightly less than  the apparent horizon, which has an area
 $4\pi H^{-2}$.  The corresponding  bound on entropy   is
 $S_{\cal V}<\pi  m_{P}^2 H^{-2}$, where $m_P$ is the Planck mass. 
 The    spacelike 3-volume enclosed by this surface, on the same spacelike hypersurface  used to describe the inflationary modes, has a proper 3-volume   $V_H=(4\pi/3)H^{-3}$. Since it is also entirely enclosed by the null surface ${\cal V}$, its entropy  is also bounded by  $S(V_H)<   \pi m_P^2 H^{-2}$. In larger 3-volumes $V>V_H$, the covariant-bound construction cannot be applied (since the inward directed light sheets have increasing surface areas); therefore,  we adopt the  conservative assumption that  entropy on larger scales is as usual an extensive quantity proportional to  $V$,   bounded by $S(V)<(V/V_H) \pi m_P^2 H^{-2}$.

No matter how physically  large a comoving volume eventually becomes, the  frozen  information contained in large-scale correlations observable at late times--- that is, all the information accessible to   classical observers, including information in any measurable quantities such as $u$ and $\vec k$--- must originate within the bounded volume $V_H$. The information per comoving volume observed in modes of any scale  cannot exceed the entropy bound per comoving volume corresponding to the time when they decohere.
This constraint sets an upper bound on  the  information in all low-$k$ modes that have frozen out, and ultimately on the density of information in classical modes. 

For large spatial volumes, classical entropy, defined as the logarithm of the number of states of a statistically uniform system,  is proportional to 3-volume in $\vec x$ space times 3- volume in $\vec k$ space, and is independent of $a$, i.e. it is conserved by the expansion.  (The 3-density of independent modes in $\vec k$ however increases as $\vec x$ volume at a given time.)  Let ${\cal I}(k')$ denote the mean frozen  information dimensionless density, per space volume times wavenumber volume, at time $t_{k'}$, in inflaton modes with $k<k'$.
To respect the   entropy bound per   3-volume  $V$,   
\begin{equation}
{\cal I}(k')   V({k'}) [ V/a^3] =   (V/V_H){\cal F} \pi m_P^2 H^{-2},
\end{equation}
where $V({k'})= 4\pi a^3 H^3/3$,  
 and ${\cal F}<1$ denotes the fraction of the covariant bound on entropy
 carried by the information in the frozen field modes with $k<k'$ at time $t_{k'}$.
 The information in the frozen mode correlations does not change after $t_{k'}$, so
the dimensionless information density (per space volume times wavenumber volume) in classical perturbations at late times is
 \begin{equation}\label{eqn:ibound}
{\cal I}    =   (9/16\pi) {\cal F} m_P^2 H^{-2}.
\end{equation}
 Equation (\ref{eqn:ibound}) is the main result of this paper. 
  It expresses the mean density  in phase space  of classical information, in terms of  parameters $H$ and $\cal F$ characterizing inflation and quantum gravity, frozen into the metric during an approximately scale-free period of inflation.

It is worth commenting that this quantity is not the same as some other measures of information in the fluctuations, such as coarse-grained entropy\cite{Kiefer:1999sj,Keski-Vakkuri:2003vj}.  The dimensionless number $\cal I$ refers to a truncated Hilbert space dimension; it represents  the logarithm of   the number of possible different frozen  field configurations, per wavenumber volume times spatial volume, representing all the different possible  classical configurations of the final spacetime metric. Thus if the covariant entropy bound applies to fields during inflation, $e^{\cal I K}$ represents  the number of possible classical observational outcomes in a  phase space volume $\cal K$.
 
\section{Physical interpretation and observability}

The holographic information bound  implies a finite bound on the number of possible observable values for  the amplitudes and   wavevectors of the classical perturbation modes. The density of all of these observables taken together cannot exceed the information bound. This places a new constraint on the kinds of classical distributions that can be realized.  It differs from the field-theory prediction that the amplitudes and phases of classical modes are  continuous random variables \cite{Maldacena:2002vr,bartolo}.  

The observability of the holographic correlations depends on the numerical value of $\cal I$.  In the general case, if ${\cal I}>>1$, it will be difficult to design realistic experiments capable of detecting this discreteness by searching for  generic nongaussian features of the fluctuations. Since $m_P^2/H^2$ is at least of the order of $10^6$, and possibly much larger than that, the effect may never be observable.

On the most optimistic scenario, if  ${\cal F}$  is small enough that $\cal I$ is of  order unity (that is,   ${\cal F}\approx H^2/m_P^2$),    holographic correlations on the current Hubble scale might explain the statistical anomalies already observed in the large angle anisotropy data\cite{deOliveira-Costa:2003pu,Hogan:2003mq}.  
If these oddities indeed reflect the holographic information bound    (rather than simple chance or, say, a customized inflation scenario, spectral discreteness due to  nontrivial topology, or other new physics\cite{Efstathiou:2003tv,Cline:2003ve,Feng:2003zu,Bastero-Gil:2003bv,Contaldi:2003zv,Tsujikawa:2003gh,Luminet:2003dx,Cornish:2003db}),  then similar correlations    are predicted also to   appear  on smaller scales.

In some models, the  information limit may also become manifest even for ${\cal F}\approx 1$.  Consider a simple model where information is encoded in a universal, fundamental scale-invariant spectrum,  with  ${\cal I}$ discrete modes in each  $\vec k$-space volume having a radius spanning a factor of $\log |k|$.   Future experiments of various kinds have enough dynamic range to detect a discrete spectrum in this situation even if ${\cal I}>>1$  \cite{Hogan:2003mq}. 
 For example, a complete galaxy catalog on the Hubble scale, within the capability of survey instruments such as the Large Synoptic Survey Telescope (LSST), will   allow independent measurement of about $10^6$ independent spatial ``pixels'' (or about the same number of independent linear plane wave modes), preserving the initial phase and amplitude information from inflation, and reaching ${\cal I}\approx 10^6$.  An advanced successor to the Laser Interferometer Space Antenna (LISA) may eventually reach the sensitivity needed to detect inflationary gravitational waves directly at 1 Hz; its
 frequency resolution will reach ${\cal I}\approx 10^8$. 
 
 Concrete, realistic predictions for such experiments require a definite holographic model of   interaction of  spacetime quanta with inflaton or graviton modes.
 In some models, the possibility in principle of discovering  the nature of the  holographic pixelation, or of setting constraints on the new physics embedded in  $\cal F$, motivates surveys  allowing detailed analysis  of spatial and temporal patterns with high sensitivity and dynamic range.

\begin{acknowledgements}
I am grateful to J.  Bardeen, R. Bousso, D. Kaplan, A. Karch, A. Nelson, D. Polarski, P. Steinhardt,  M. Strassler, L. Susskind, and  M. Turner  for useful comments and discussions, and to Fermilab for hospitality.
This work was supported by NSF grant AST-0098557 at the University of
Washington.

\end{acknowledgements}

\end{document}